\documentclass[aps,preprint,floatfix,nofootinbib,showpacs]{revtex4-1}

\pdfoutput=1
\usepackage{graphicx,color,float}
\usepackage{hyperref}
\usepackage{multirow}
\usepackage{verbatim}
\usepackage{longtable}
 \usepackage{amsmath}
\usepackage{amsfonts}
\usepackage{amssymb}
\usepackage{rotating}
\usepackage[dvipsnames]{xcolor}

\newcommand{\lsim}{\raisebox{-0.13cm}{~\shortstack{$<$ \\[-0.07cm] $\sim$}}~}

\begin{document}

{\small
\begin{flushright}
IUEP-HEP-20-01
\end{flushright} }

\title{
Measuring the trilinear Higgs boson self--coupling 
at the 100 TeV hadron collider via multivariate analysis
}

\def\slash#1{#1\!\!/}

\renewcommand{\thefootnote}{\arabic{footnote}}

\author{
Jubin Park$^{1,2}$,
Jung Chang$^{2}$, Kingman Cheung$^{3,4,5}$, 
and Jae Sik Lee$^{2,1}$}
\affiliation{
$^1$ IUEP, Chonnam National University, Gwangju 61186, Korea \\
$^2$ Department of Physics, Chonnam National University,
Gwangju 61186, Korea\\
$^3$ Physics Division, National Center for Theoretical Sciences,
Hsinchu, Taiwan \\
$^4$ Division of Quantum Phases and Devices, School of Physics, 
Konkuk University, Seoul 143-701, Republic of Korea \\
$^5$ Department of Physics, National Tsing Hua University,
Hsinchu 300, Taiwan 
}
\date{September 11, 2020}

\begin{abstract}
We perform a multivariate analysis of Higgs-pair production 
via the decay channel  $HH \to b\bar b \gamma\gamma$ at the
future 100 TeV $pp$ collider to determine
the trilinear Higgs self--coupling (THSC) $\lambda_{3H}$, which takes
the value of 1 in the standard model.
We consider all known background processes.
For the signal we adopt the most recent event generator of
{\tt POWHEG-BOX-V2} to exploit the NLO distributions 
for Toolkit for Multivariate Data Analysis (TMVA).
%
Through the technique of Boosted Decision Tree (BDT) analysis trained
for $\lambda_{3H}=1$,
compared to the the conventional cut-and-count approach,
the signal-to-background ratio improves tremendously 
from about $1/10$ to $1$
and the significance can reach up to $20.5$ with a luminosity of 3 ab$^{-1}$
without including systematic uncertainties.
In addition, by implementing a likelihood fitting of the signal-plus-background
$M_{\gamma\gamma b b}$ distribution with optimized bin sizes, 
it is possible to determine
the THSC with the precision of 7.5\% at 68\% CL
even at the early stage of 100 TeV hadron collider with 3 ab$^{-1}$.
\end{abstract}

\maketitle

\section{Introduction} 
Since the discovery of the 125 GeV Higgs boson in 2012 at the
LHC \cite{discovery},
we have been looking for a clear signal or 
even a hint of new physics beyond the Standard Model (SM) but without much success.
Moreover, after completing the Runs I and II at the LHC, 
it turns out that the 125 GeV Higgs boson is best described as 
the SM Higgs boson \cite{higgcision},
although there is an upward trend in the overall signal
strength \cite{Cheung:2018ave}.
Under this situation,  one of the most solid avenues to explore for new physics
is to measure the Higgs potential which could be significantly different from
that of the SM.

Higgs-boson pair production at the high-luminosity and/or high-energy
hadron colliders provides a very useful way to probe the Higgs potential
via the investigation of the trilinear Higgs self-coupling 
(THSC)~\cite{dihiggs1,dihiggs2,dihiggs3}.
The specific decay modes considered are: $b\bar b b\bar b$~\cite{bbbb},
$b\bar b\gamma\gamma$~\cite{bbaa,bbaa_atlas17},
$b\bar b\tau^+\tau^-$~\cite{bbtata},
$b\bar b W^+W^-$~\cite{bbWW}, and some combinations
of these channels~\cite{decaycombined,comb_atlas18}.
Higgs-boson pair production also has been vastly studied in 
models beyond the SM~\cite{dihiggs_np}.

The current limits on the THSC in units of $\lambda_{3H}$, which takes
the value of 1 in the SM, are
$-5.0 < \lambda_{3H} < 12$  from ATLAS~\cite{Aad:2019uzh}
and $-11.8 < \lambda_{3H} < 18.8$  from CMS~\cite{Sirunyan:2018two}
at 95\% confidence level (CL). 
At the high-luminosity option of the LHC running at 14 TeV (HL-LHC)
with an integrated luminosity of 3 ab$^{-1}$, 
a combined ATLAS and CMS projection of the 68\% CL interval 
is $0.57 < \lambda_{3H} < 1.5$ 
without including systematic uncertainties~\cite{Cepeda:2019klc}.
On the other hand, at the International Linear Collider (ILC) operated at
1 TeV can reach the precision of 10\% at 68\%  CL with 
an integrated luminosity of 8 ab$^{-1}$~\cite{Fujii:2015jha,Braathen:2019zoh}.

\bigskip

In this work, we perform a multivariate analysis of Higgs-pair production 
in $HH \to b\bar b \gamma\gamma$ channel at the 
100 TeV hadron collider.
In our previous work,   
based on the conventional cut-and-count analysis,
it was shown that the THSC can be measured with about 20\% accuracy
at the SM value with a luminosity of 3 ab$^{-1}$~\cite{Chang:2018uwu}.
%
In this Letter, with the use of the BDT method 
closely following Ref.~\cite{Chang:2019ncg},
we show that the THSC can be measured with 
a precision of $7.5\%$ at 68\% CL at the 100 TeV hadron collider
assuming 3 ab$^{-1}$ luminosity,
which is superior to the accuracy expected
at the 1 TeV ILC even with 8 ab$^{-1}$.

\section{Event generation and TMVA analysis}
The Higgs bosons in the signal event samples
are generated on-shell with a zero width by
\texttt{POWHEG-BOX-V2}~\cite{Heinrich:2017kxx,Heinrich:2019bkc} with
the damping factor $\mathtt{hdamp}$ set to the default value of 250
to limit the amount of hard radiation. 
This code provides NLO distributions matched to a parton shower 
taking account of the full top-quark mass dependence. 
The signal cross section at NNLO order in QCD is calculated according to
$\sigma^{\rm NNLO} ( \lambda_{3H} )=
K^{\rm NNLO / NLO}_{\rm SM}\, \sigma^{\rm NLO}  (\lambda_{3H}) \;$
using $\sigma^{\rm NLO}(\lambda_{3H})$ from {\tt POWHEG-BOX-V2}
and  $K^{\rm NNLO/NLO}_{\rm SM}=1.067$~\cite{Grazzini:2018bsd}
\footnote{
According to the recent N3LO calculations~\cite{n3lo}, 
the signal cross section is further enhanced by the amount of $2.7$\% which 
would hardly affect our conclusion, or rather strengthen our results.}
in the FT
approximation in which the full top-quark mass dependence is considered only
in the real radiation while the Born improved Higgs Effective Field Theory
is taken in the virtual part.
And then, 
the \texttt{MadSpin} code \cite{Artoisenet:2012st} is used for the decay of
both Higgs bosons into two bottom quarks and two photons.

For generation and simulation of backgrounds,
we closely follow Ref.~\cite{Chang:2018uwu}
\footnote{Specifically, the multi-variate MV1 $b$-tagging algorithm 
with $\epsilon_b=0.75$ is taken together with $P_{c\to b}=0.1$, 
$P_{j\to b}=0.01$, and $P_{j\to\gamma}=1.35\times 10^{-3}$~\cite{Contino:2016spe}.}, 
except for the use
of the post-LHC PDF set of {\tt CT14LO}~\cite{Dulat:2015mca} 
for non-resonant backgrounds.
Furthermore,
for the two main non-resonant backgrounds of $b\bar b\gamma\gamma$ and $c\bar
c\gamma\gamma$,
we use the merged cross sections and distributions
by MLM matching~\cite{Mangano:2006rw,Alwall:2007fs}
with {\bf xqcut} and $Q_{\rm cut}$ set to 20 GeV and 30 GeV, respectively.
For the remaining non-resonant backgrounds,
we are using the cross sections and distributions obtained by applying
the generator-level cuts as adopted in Ref.~\cite{bbaa_atlas17,comb_atlas18}
which might provide more reliable and conservative estimation of the non-resonant
backgrounds
containing light jets~\cite{Chang:2018uwu}.

For parton showering and hadronization,
\texttt{PYTHIA8}~\cite{Sjostrand:2014zea} is used both for
signal and backgrounds.
Finally, fast-detector simulation and analysis are
performed using \texttt{Delphes3}~\cite{deFavereau:2013fsa}
with the \texttt{Delphes-FCC} template.

All the signal and backgrounds are summarized in Table~\ref{tab:ParticleList_100TeV},
together with information of the corresponding event
generator, the cross section times the branching ratio and 
the order in QCD, and the Parton Distribution Function (PDF) used.

\begin{table}[th!]
\centering
\caption{\small 
Monte Carlo samples used in Higgs-pair production
analysis $H(\rightarrow b\bar{b})H(\rightarrow \gamma\gamma)$, and
the corresponding codes for the matrix-element generation. 
$\tt{PYTHIA8}$ is used for parton showering and hadronization. 
We refer to Ref.~\cite{Alwall:2014hca} for $\mathtt{MG5\_aMC@NLO}$.
}
\vspace{3mm}
\label{tab:ParticleList_100TeV}
\begin{tabular}{  c  c  c  c  c  c }
\hline
\multicolumn{6}{c}{Signal} \\
\hline
\multicolumn{2}{c}{Signal process} & Generator &
$\sigma \cdot BR$ [fb] & Order  & PDF used  \\
&&&& in QCD &   \\
\hline
\multicolumn{2}{c}{$gg \to HH \to b\bar b \gamma\gamma$} &
 {\tt POWHEG-BOX-V2} & 3.25
 & NNLO &  PDF4LHC15$\_$nlo \\ 
\hline
\hline
\multicolumn{6}{c}{Backgrounds} \\
\hline
Background(BG)  & Process  & Generator & $\sigma\cdot BR$~[fb] &  Order  &
PDF used\\
&&&& in QCD&   \\
\hline
\multirow{4}{*}{}
& $ggH(\rightarrow \gamma\gamma)$  &  $\mathtt{POWHEG-BOX}$
  & $1.82 \times 10^{3}$ & $\mathrm{NNNLO}$ & $\mathtt{CT10}$ \\ \cline{2-5}
Single-Higgs  & $t \bar{t} H(\rightarrow \gamma\gamma)$  &
$\mathtt{PYTHIA8}$  & $7.29\times 10^1$ & NLO & \\ \cline{2-5}
 associated BG             &  $ZH(\rightarrow \gamma\gamma)$ &
$\mathtt{PYTHIA8}$  & $2.54\times 10^1$ & NNLO &\\ \cline{2-5}
                  & $b\bar{b}H(\rightarrow \gamma\gamma)$ &
$\mathtt{PYTHIA8}$ & $1.96\times 10^1$ & NNLO(5FS)  &\\ \hline
\multirow{7}{*}{Non-resonant BG} 
 & $b\bar{b} \gamma\gamma$ & $\mathtt{MG5\_aMC@NLO}$ & $2.28 \times 10^3$  &  LO & CT14LO \\  \cline{2-5}
 &  $c\bar{c} \gamma\gamma$ & $\mathtt{MG5\_aMC@NLO}$ &  $1.92 \times 10^4$  & LO &  MLM~\cite{Mangano:2006rw,Alwall:2007fs}  \\ \cline{2-6}
 &  $jj\gamma\gamma$ & $\mathtt{MG5\_aMC@NLO}$& $4.20 \times 10^5$  & LO &  \\ \cline{2-5}
 &  $b\bar{b}j\gamma$ & $\mathtt{MG5\_aMC@NLO}$ & $0.96 \times 10^7$ & LO & \\ \cline{2-5}
 &  $c\bar{c}j\gamma$ & $\mathtt{MG5\_aMC@NLO}$& $3.19 \times 10^7$ & LO &  CT14LO \\ \cline{2-5}
 &  $b\bar{b}jj$  & $\mathtt{MG5\_aMC@NLO}$ & $1.00 \times 10^{10}$ & LO &  
Refs.~\cite{bbaa_atlas17,comb_atlas18,Chang:2018uwu}  \\ \cline{2-5}
 & $Z(\rightarrow b\bar{b})\gamma\gamma$ &
$\mathtt{MG5\_aMC@NLO}$ &  $7.87 \times 10^1)$ & LO & \\ \hline
\multirow{2}{*}{$t\bar{t}$ and $t\bar{t}\gamma$ BG} 
& $t\bar{t}$ & $\mathtt{MG5\_aMC@NLO}$  & $1.76 \times 10^7$ & NLO &  $\mathtt{CT10}$  \\ 
\cline{2-6}
  ($\geq 1$ lepton)   & $t\bar{t}\gamma$
 & $\mathtt{MG5\_aMC@NLO}$  &
$4.18 \times 10^4$ & NLO & $\mathtt{CTEQ6L1}$ \\
\hline
\end{tabular}
\end{table}
%

%
%
%
\begin{table}[th!]
\caption{
Sequence of event selection criteria applied in this analysis.}
\label{tab:event_selection}
\vspace{3mm}
  \begin{tabular}{|c|l| }
  \hline
    Sequence &~ Event Selection Criteria at the 100 TeV hadron collider \\
    \hline
    \hline
    1 &~ Di-photon trigger condition,
$\geq $ 2 isolated photons with $P_T > 30$ GeV, $|\eta| < 5$
\\
    \hline
    2 &~ $\geq $ 2 isolated photons with $P_T > 40$ GeV,
$|\eta| < 3$, $\Delta R_{j\gamma\,,\gamma\gamma} > 0.4$ \\
    \hline
    3 &~ $\geq$ 2 jets identified as b-jets with leading(subleading) $P_T > 50(40)$ GeV, 
$|\eta|<3$, $\Delta R_{bb} > 0.4$ \\
    \hline
    4 &~ Events are required to contain $\le 5$ jets with
 $P_T >40$ GeV within $|\eta|<5$ \\
\hline
    5 &~ No isolated leptons with $P_T > 40$ GeV, $|\eta| <3$ \\
    \hline
    6 &~ TMVA analysis  \\
    \hline
    \end{tabular}
\end{table}
A multivariate analysis is performed using TMVA~\cite{TMVA2007}
with {\tt ROOTv6.18}~\cite{ROOT}. 
After applying a sequence of event selections
as in Table~\ref{tab:event_selection},
we choose the following 8 kinematic variables for TMVA:
\begin{equation}
M_{bb}\,, \ \ P_{T}^{bb}\,, \ \ \Delta R_{bb}\,; \ \
M_{\gamma\gamma}\,, \ \ P_{T}^{\gamma\gamma}\,, \ \ \Delta R_{\gamma\gamma}\,; \ \
M_{\gamma\gamma bb}\,, \ \ \Delta R_{\gamma b}\,. \nonumber
\end{equation}
The judicious choice of the two photons or two $b$ quarks for the above TMVA variables
has been made as in ~\cite{Chang:2019ncg}.
We also refer to Ref.~\cite{Chang:2019ncg} for the details of our TMVA setup and
analysis. And we choose  BDT for our analysis since
the BDT-related methods show higher performance
with better signal efficiency and stronger background rejection.

%
\begin{figure}[t!]
\centering
\includegraphics[width=3.2in,height=2.9in]{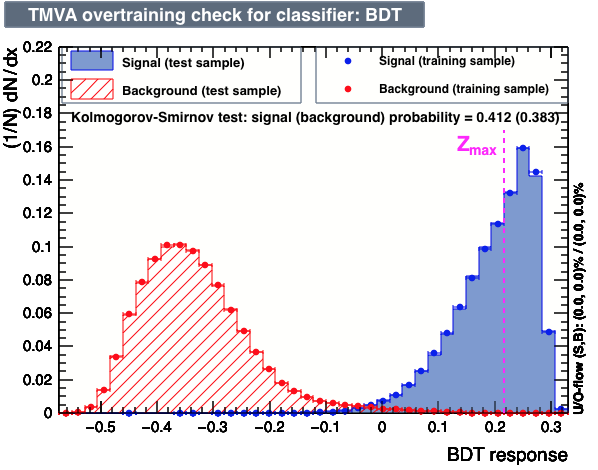}
\includegraphics[width=3.2in,height=2.9in]{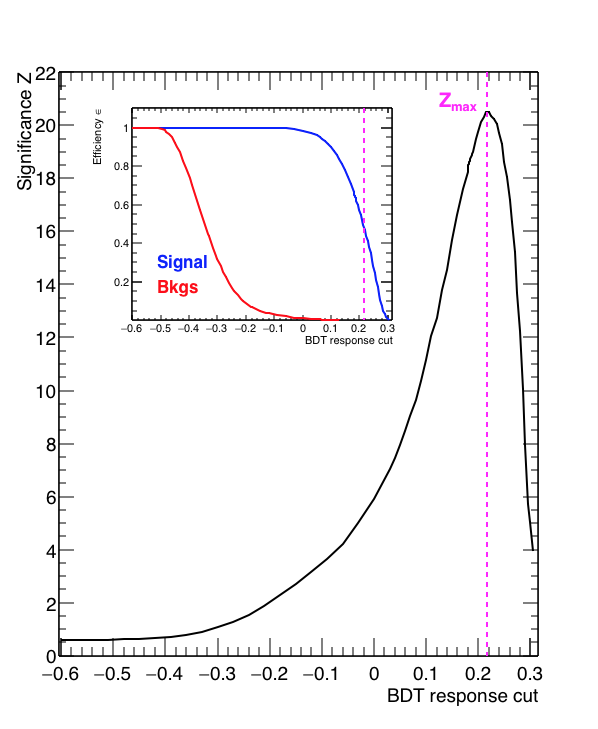}
\caption{
(Left) 
Normalized SM BDT responses for test (histogram) and training (dots with error bars) 
samples. BDT responses for signal (blue) and background (red) samples,
which mostly populate
in the regions with positive and negative BDT response, respectively.
(Right)
Signal and background efficiencies (inset) and
significance $Z$ as functions of BDT response cut.
BDT$_{\rm SM}$ is used.
The vertical lines show the position of 
the optimal cut on the BDT response which maximizes the significance.
}
\label{fig:BDT_over}
\end{figure}
\section{Results}
In the left panel of Fig.~\ref{fig:BDT_over}, we show the BDT responses obtained using
BDT trained for $\lambda_{3H}=1$ which is dubbed as BDT$_{\rm SM}$.
By validating the BDT distributions for the training sample (dots with error bars)
with those for the test sample (histogram), we check that BDT$_{\rm SM}$ is not 
overtrained. 
In the right panel of Fig.~\ref{fig:BDT_over}, using BDT$_{\rm SM}$,
we show the behavior of signal and background efficiencies (inset)
and significance 
$ Z = \sqrt{ 2 \cdot \left[ \left( (s+b) \cdot \ln( 1+ s/b) - s \right)  \right ] }$
with $s$ and $b$ being the numbers of signal and background events
as functions of the cut value on BDT response.
The significance can reach up to $20.50$ when the BDT response is cut
at $0.216$, at which, the signal and background efficiencies are
$0.48$ and $1.58\times 10^{-4}$, respectively.
We denote by vertical lines the positions of the optimal cut on 
the BDT response which maximizes the significance.

%
\begin{table}[t!]
\caption{\small
Expected number of signal and background events at the 100 TeV hadron collider 
assuming 3 ab$^{-1}$ using BDT$_{\rm SM}$ with the BDT response cut of $0.216$.
See text for explanation.
}
\vspace{-1mm}
\label{tab:l3h1_opt}
\begin{center}
\begin{tabular}{| l || r | r || r || r |}
\hline
&\multicolumn{3}{c||}{Expected yields $(3~ \mathrm{ab}^{-1})$}
& \\ \hline
Signal and Backgrounds & ~Pre-Selection & ~BDT$_{\rm SM}$ &  ~Cut-and & Eff. Lumi.  \\ 
 &  & &  -Count~~ & (ab$^{-1}$)~~  \\ 
\hline
$H(b\,\bar{b})\,H(\gamma\,\gamma)$, $\lambda_{3H} = -3$ & 7253.98	&2408.37 &	3400.08&10.7 \\
$H(b\,\bar{b})\,H(\gamma\,\gamma)$, $\lambda_{3H} = 0$ & 2072.09	&	902.49&	1146.21&44.5 \\
$\mathbf{H(b\,\bar{b})\,H(\gamma\,\gamma)}$, $\mathbf{\lambda_{3H}= 1}$ & \textbf{1124.48} & \textbf{548.02} & \textbf{673.29}  &\textbf{615} \\
$H(b\,\bar{b})\,H(\gamma\,\gamma)$, $\lambda_{3H} = 5$ & 1480.24	&251.13 &	439.29&40.9 \\ \hline
$gg\,H(\gamma\,\gamma)$ &5827.41	&255.86 &	875.71&17.0 \\
$t\,\bar{t}\,H(\gamma\,\gamma)$ &11371.21	&145.88 &	868.73&13.2 \\
$Z\,H(\gamma\,\gamma)$ & 593.29	&38.88 &	168.86&39.4 \\
$b\,\bar{b}\,H(\gamma\,\gamma)$ & 205.45	&2.59 &	9.82&51.0 \\ \hline
$b\,\bar{b}\,\gamma\,\gamma$&183493.56	&	55.01&	336.49 &19.2 \\
$c\,\bar{c}\,\gamma\,\gamma$ & 66600.78	&0.00 &	54.66&0.11 \\
$j\,j\,\gamma\,\gamma$ &14182.56	&2.52 &	25.20&2.38 \\
$b\,\bar{b}\,j\,\gamma$ & 1228956.91	&	38.53&	1176.93&3.74 \\
$c\,\bar{c}\,j\,\gamma$ &208285.83	&0.00 &	187.92&0.26 \\
$b\,\bar{b}\,j\,j$ & 1622778.23	& 0.00 &	2231.08&0.19 \\
$Z(b\,\bar{b})\,\gamma\gamma$ & 4540.20	&4.72 &	45.33&12.7 \\ \hline
$t\,\bar{t}$~($\geq$ 1 leptons)& 78490.03 &0.00&56.93&~$11.5+3.68$ \\
$t\,\bar{t}\,\gamma$~($\geq$ 1 leptons)& 74885.54 &  9.09&105.16&~$8.69+2.07$ \\ \hline
Total Background&3500211.00 &	553.09&	6142.83& \\ \hline\hline
Significance $Z$, $\lambda_{3H} = 1$ && \textbf{20.50} & \textbf{8.44}& \\ \hline
\end{tabular}
\end{center}
\end{table}
In Table~\ref{tab:l3h1_opt}, we present the
expected number of signal and background events at the 100 TeV hadron 
collider assuming 3 ab$^{-1}$ using BDT$_{\rm SM}$ with the BDT response cut of $0.216$.
We show the four representative values of $\lambda_{3H}$ for signal
and the backgrounds are separated into three categories.
For comparisons, we also show the results obtained using the cut-and-count
analysis~\cite{Chang:2018uwu}.
In the last column, we additionally present the effective luminosity (Eff. Lumi.) for each
of signal and background samples.
In the $t\bar t$ and $t\bar t\gamma$ backgrounds,
the first (second) number is the effective luminosity
when the two top quarks decay fully (semi-) leptonically.
%
%
We find about $550$ signal and $550$
background events for $\lambda_{3H}=1$. Comparing to the results using
the cut-and-count analysis~\cite{Chang:2018uwu},
the number of signal events
decreases by only $19\%$ while the number of backgrounds by almost $90\%$, resulting in 
an increase in significance from $8.44$ to $20.50$.
Note that the composition of backgrounds changes drastically by the use of BDT.
In the cut-and-count analysis, the non-resonant background is about two times 
larger than the single-Higgs associated background. 
While, in the BDT analysis,
the single-Higgs associated background is more than four times  
larger than the non-resonant one
and $t\bar t$ associated background becomes negligible.
Note that we generate relatively smaller number of events for
the $c\bar c\gamma\gamma$, $c\bar cj\gamma$, and
$b\bar bjj$ backgrounds since we observe that they quickly decrease when
the BDT response cut approaches to the point $Z_{\rm max}$ of $0.216$
\footnote{In fact,
there are some differences in kinematic distributions among 
the non-resonant backgrounds. For example, 
the $c\bar c\gamma\gamma$ background is more populated in the
region of $\Delta R_{bb}>3$ compared to the $b\bar b\gamma\gamma$ one.}.
Specifically, the $b\bar b j j$ background vanishes
for the BDT response cut larger than $0.2$.
Otherwise, we generate enough number of events considering the assumed luminosity of
3 ab$^{-1}$.

\begin{figure}[t!]
\centering
\includegraphics[width=3.12in,height=2.4in]{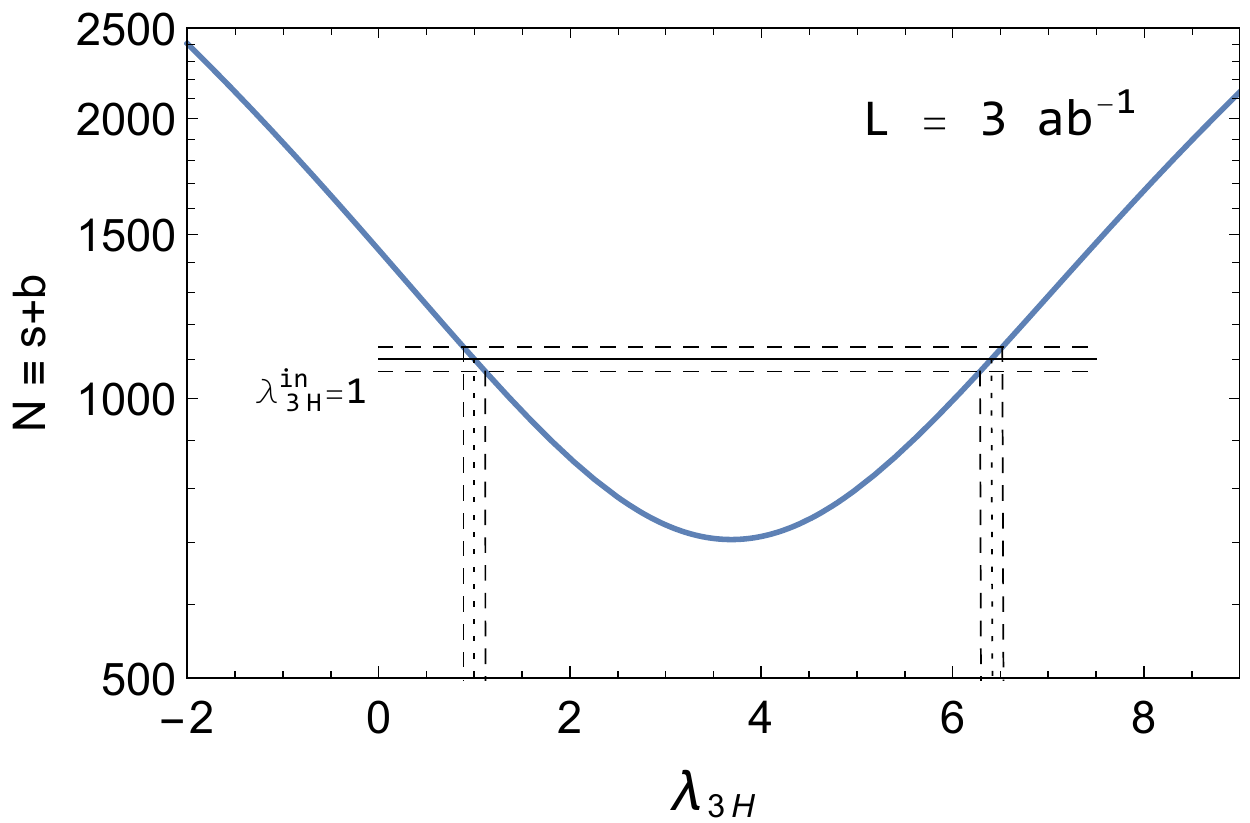}
\includegraphics[width=3.12in,height=2.4in]{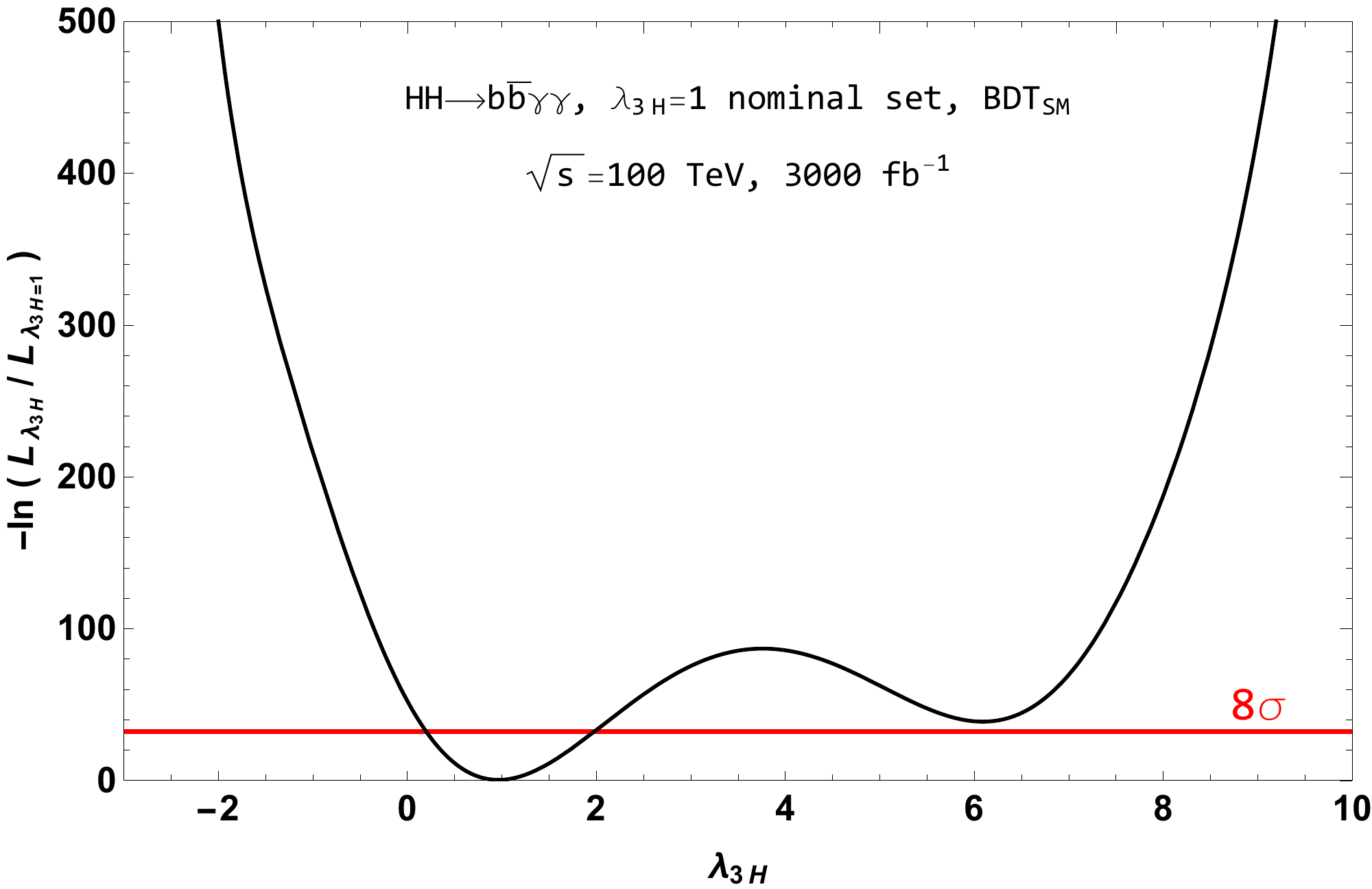}
\caption{
(Left) The total number $N =s+b$ of signal ($s$) and background ($b$) events
versus $\lambda_{3H}$ with 3 $\mathrm{ab}^{-1}$.
The horizontal solid line denotes the total number of events
obtained using the SM value of $\lambda_{3H}=1$
and the dashed lines for the statistical $1$-$\sigma$ error.
(Right) 
The relative log likelihood distribution 
for the nominal value of $\lambda_{3H}=1$
at the 100 TeV hadron collider assuming 3 ab$^{-1}$ and using
BDT$_{\rm SM}$ with the BDT response cut of $0.216$.
The distribution has been obtained by a likelihood
fitting of $M_{\gamma\gamma b b}$ distribution for each value of 
$\lambda_{3H}$.
The black solid line shows the result of a polynomial fitting
and the horizontal 
solid (red) line at $-\ln(L_{\lambda_{3H}}/L_{\lambda_{3H}=1})=32$ indicates the value
corresponding to the $8\sigma$ level.
}
\label{fig:mhh_BDTL1}
\end{figure}
First, we try to determine the THSC considering the total number of events.
As shown in the left panel of Fig.~\ref{fig:mhh_BDTL1},
we find that the THSC can be measured with about $11\%$ accuracy
at the SM value which is about two times better than the result
based on the conventional cut-and-count analysis~\cite{Chang:2018uwu}.
However, there is a second solution around $\lambda_{3H}=6.5$.
To lift up the two-fold ambiguity,
we implement a likelihood fitting of the signal-plus-background
$M_{\gamma\gamma bb}$ distribution
and find the second solution is ruled out 
 by more than $8\sigma$ confidence,
see the right panel of Fig.~\ref{fig:mhh_BDTL1}.

\begin{figure}
\centering
\includegraphics[width=3.12in,height=2.4in]{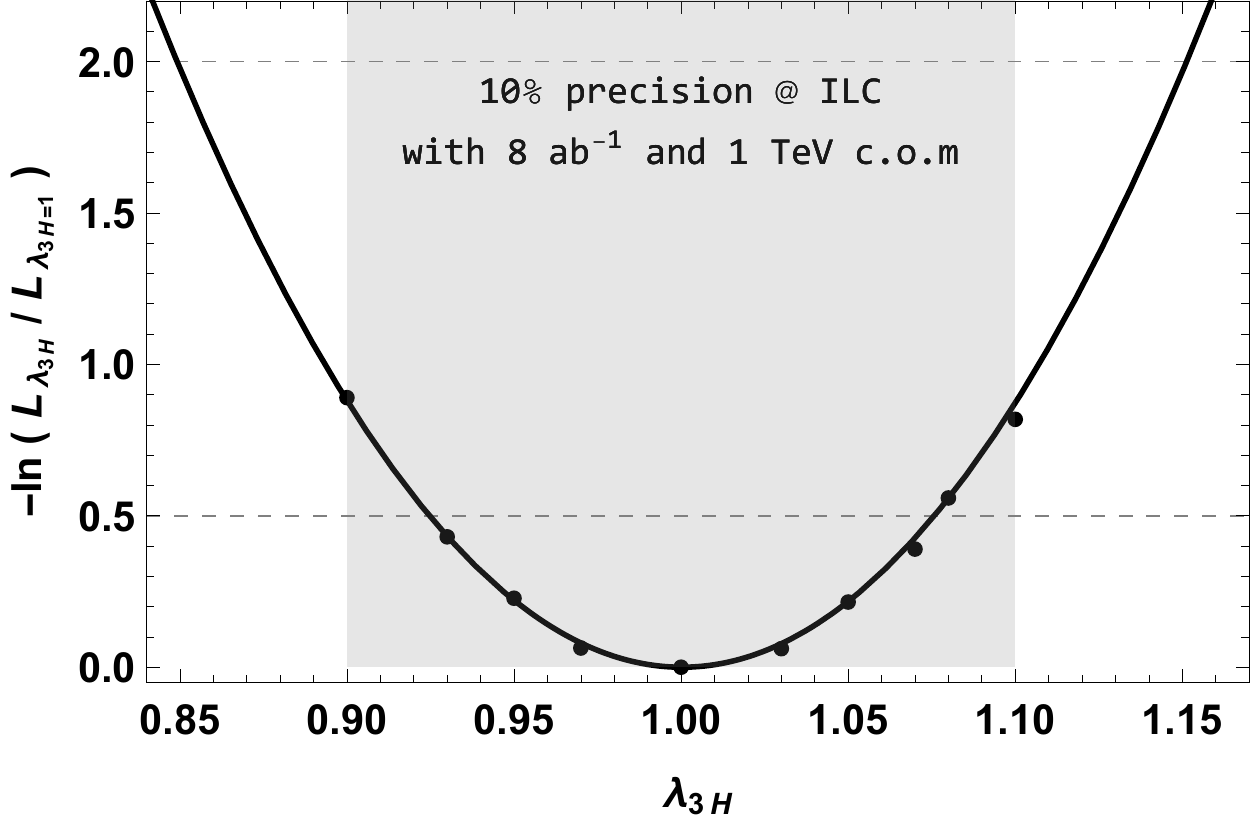}
\includegraphics[width=3.12in,height=2.4in]{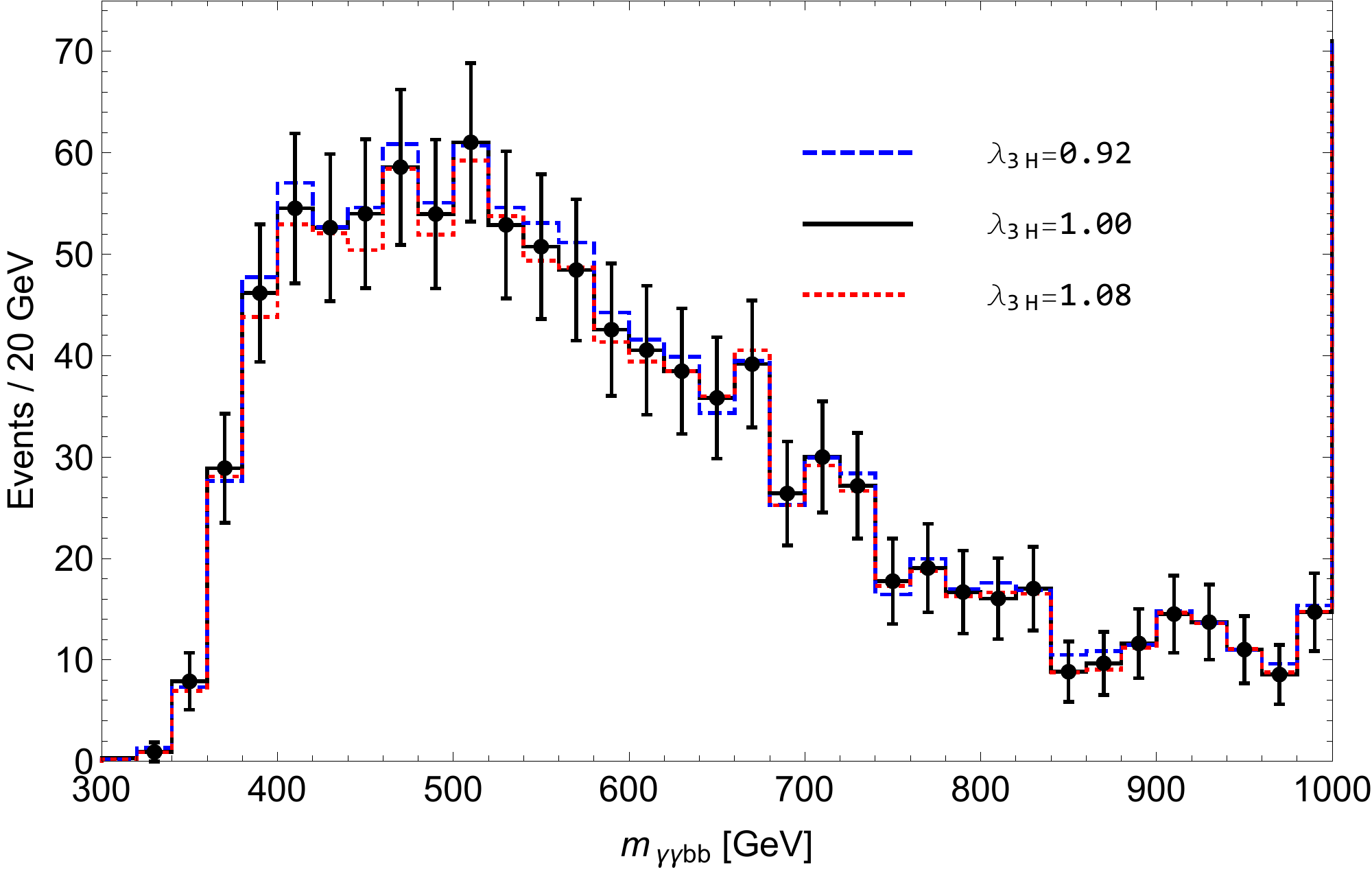}
\caption{
(Left)
The relative log likelihood distribution 
for the nominal value of $\lambda_{3H}=1$
at the 100 TeV hadron collider with 3 ab$^{-1}$.
The black circles are the values obtained by a likelihood
fitting of $M_{\gamma\gamma b b}$ distributions 
using BDT$_{\rm SM}$ with the BDT response cut of $0.216$.
The black solid line shows the result of a polynomial fitting
and the thin dashed line at $0.5\,(2.0)$ indicates the value corresponding to
a $1\sigma\,(2\sigma)$ CI.
The shaded region shows the $1\sigma$ CI expected at the ILC
at 1 TeV with 8 ab$^{-1}$.
(Right)
The SM $M_{\gamma\gamma bb}$ distribution (solid line with dots with $1\sigma$ error bars)
and those for $\lambda_{3H}=0.92$ and $1.08$ (dashed lines).
}
\label{fig:result}
\end{figure}

  To improve the sensitivity of the THSC around the SM value 
  and to tame the statistical fluctuation due to the limited
  size of the MC samples, we repeat the likelihood fitting of
  $M_{\gamma\gamma bb}$ distribution by optimizing the bin size
  between $1/20\,$GeV and $1/60\,$GeV.
Finally, we find that
the THSC can be determined with a precision of $7.5\%$ at 68\% CL
as shown in the left panel of Fig.~\ref{fig:result}. 
In the right panel of Fig.~\ref{fig:result}, $M_{\gamma\gamma bb}$ distributions
are shown for
the THSC at the SM value and for the two values
deviated by $1\sigma$.

\begin{figure}
\centering
\includegraphics[width=4.12in,height=3.17in]{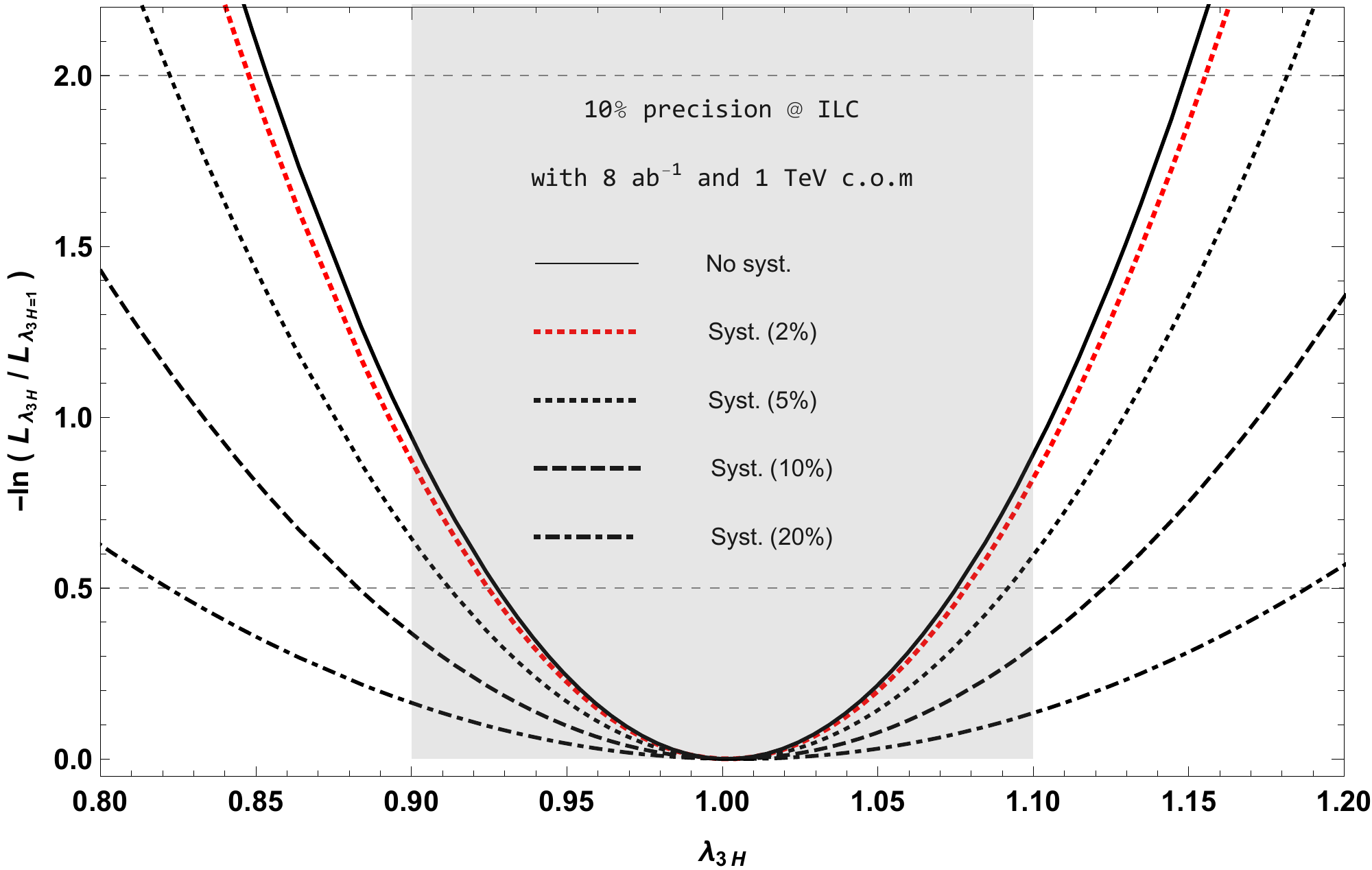}
\caption{
The same as in the left panel of Fig.~\ref{fig:result} while taking
$\sigma_b/b = 0$ (solid), $0.02$ (red dotted)
$0.05$ (black dotted), $0.1$ (dashed), and $0.2$ (dash-dotted).
}
\label{fig:sys}
\end{figure}

By now, we have considered only the statistical uncertainties which
may eventually dominate the total uncertainties.
Before concluding, we would like to discuss the effects of 
systematic uncertainties which could be important
at the early stage of 100 TeV hadron collider.
The systematic uncertainties might be taken into account by 
considering the variance of background $\sigma_b^2$~\cite{Cowan:2010js}.
In this case, the error due to systematic uncertainties is proportional to the number
of background or $\sigma_b \propto b$.
%
We find that the THSC precision of $7.5$\%$-18$\%  at 68\% CL
while varying $\sigma_b/b$ between $0$
and $0.2$, see Fig.~\ref{fig:sys}
\footnote{Incidentally, by measuring only the total number of events,
the precision becomes worse to $11$\%$-30$\%.}.

Finally, before we end this section, in Table~\ref{ranking},
we show the relative importance of the variables
that we employed in this BDT analysis.
We observe that
the two most important variables are $\Delta R_{bb}$ and $\Delta R_{\gamma\gamma}$,
which is consistent with our previous cut-and-count analysis
\cite{Chang:2018uwu}.

\begin{table}[h]
  \caption{\small \label{ranking}
    The ranking of the variables that we employed in this BDT analysis
    in the descending order of importance.}
\label{tab:ranking}
\vspace{0.1cm}
  \begin{ruledtabular}
    \begin{tabular}{cccccccc}
    $\Delta R_{bb}$ & $\Delta R_{\gamma\gamma}$ & $M_{\gamma\gamma}$ &
    $\Delta R_{\gamma b}$ & $P_T^{\gamma\gamma}$ & $M_{\gamma\gamma b b}$
      & $P_T^{bb}$ & $M_{bb}$ \\
      \hline
    0.163 & 0.152 & 0.150 & 0.133 & 0.110 & 0.102 & 0.096 & 0.095
    \end{tabular}
    \end{ruledtabular}
    \end{table}

\section{Conclusions:}
Higgs-pair production is one of the most useful avenue to probe the EWSB
sector. We have studied in great details, with the help of
machine learning, the sensitivity of measuring the THSC $\lambda_{3H}$
that one can expect at the 100 TeV $pp$ collider with an integrated
luminosity 3 ab$^{-1}$. With TMVA one can improve
the signal-to-background ratio for $\lambda_{3H}=1$
to $1:1$ compared with the ratio $1:10$ obtained in the conventional
cut-and-count approach.
Furthermore, the significance of such a signal jumps to $20$.

Other than determining the THSC by measuring the total number of events,
one can also improve the sensitivity and lift the two-fold degeneracy
by implementing a likelihood fitting of the signal-plus-background
$M_{\gamma\gamma b b}$ distribution with optimized bin sizes. The THSC
can be determined with a precision of 7.5\% at 68\% CL
with 3 ab$^{-1}$, which is indeed
better than the ILC running at 1 TeV with 8 ab$^{-1}$.
Extrapolating our result
conservatively, we expect that one can achieve the precision better than
$\sim 2$\% with 30 ab$^{-1}$.

\bigskip

\noindent
{\it Note added}: After the completion of our work, we learned 
a similar analysis performed considering various systematic 
uncertainties rigorously~\cite{Mangano:2020sao}.
They found the combined
precision of $2.9$\%$-5.5$\% with 30 ab$^{-1}$ at 68\% CL
which is in a good quantitative agreement with our results.

\newpage
\section*{Appendix}
\def\theequation{\Alph{section}.\arabic{equation}}
\begin{appendix}

\setcounter{equation}{0}
\section{More on the $c\bar c\gamma\gamma$, $c\bar cj\gamma$, and
$b\bar bjj$ backgrounds}
For this work,
we generate relatively smaller number of events for
the $c\bar c\gamma\gamma$, $c\bar cj\gamma$, and
$b\bar bjj$ backgrounds which may lead to underestimation
of the relevant backgrounds.

The $c\bar c\gamma\gamma$ and $c\bar cj\gamma$ backgrounds might be
negligible since, taking account of the fake rates $P_{c\to b}$ and
$P_{j\to\gamma}$, the cross sections are smaller than
that of the $b\bar b\gamma\gamma$ background by about an order of magnitude.
On the other hand, our estimation of the $b\bar bjj$ background
could be unreliable due to the limited size of the MC sample.
Here we try to estimate the background yield based on the current sample.

\begin{figure}[h!]
\centering
\includegraphics[width=3.18in]{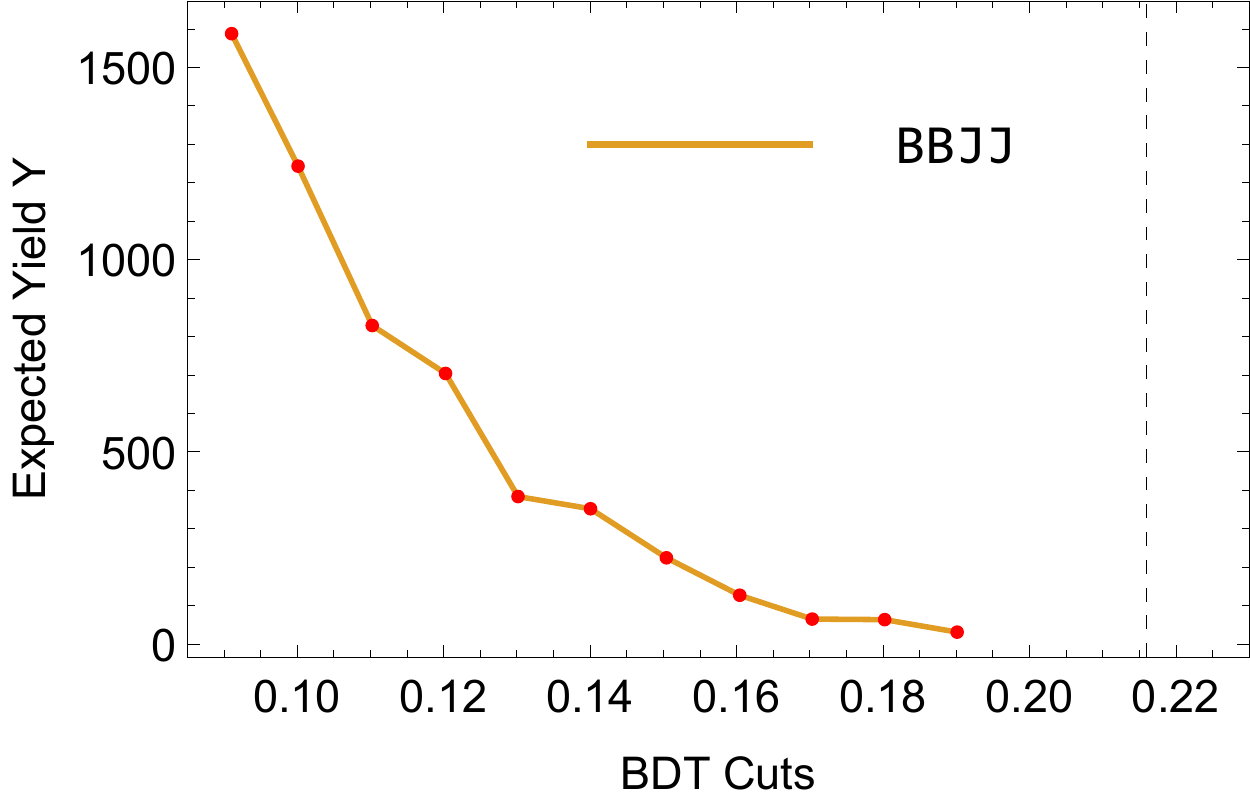}
\includegraphics[width=3.0in]{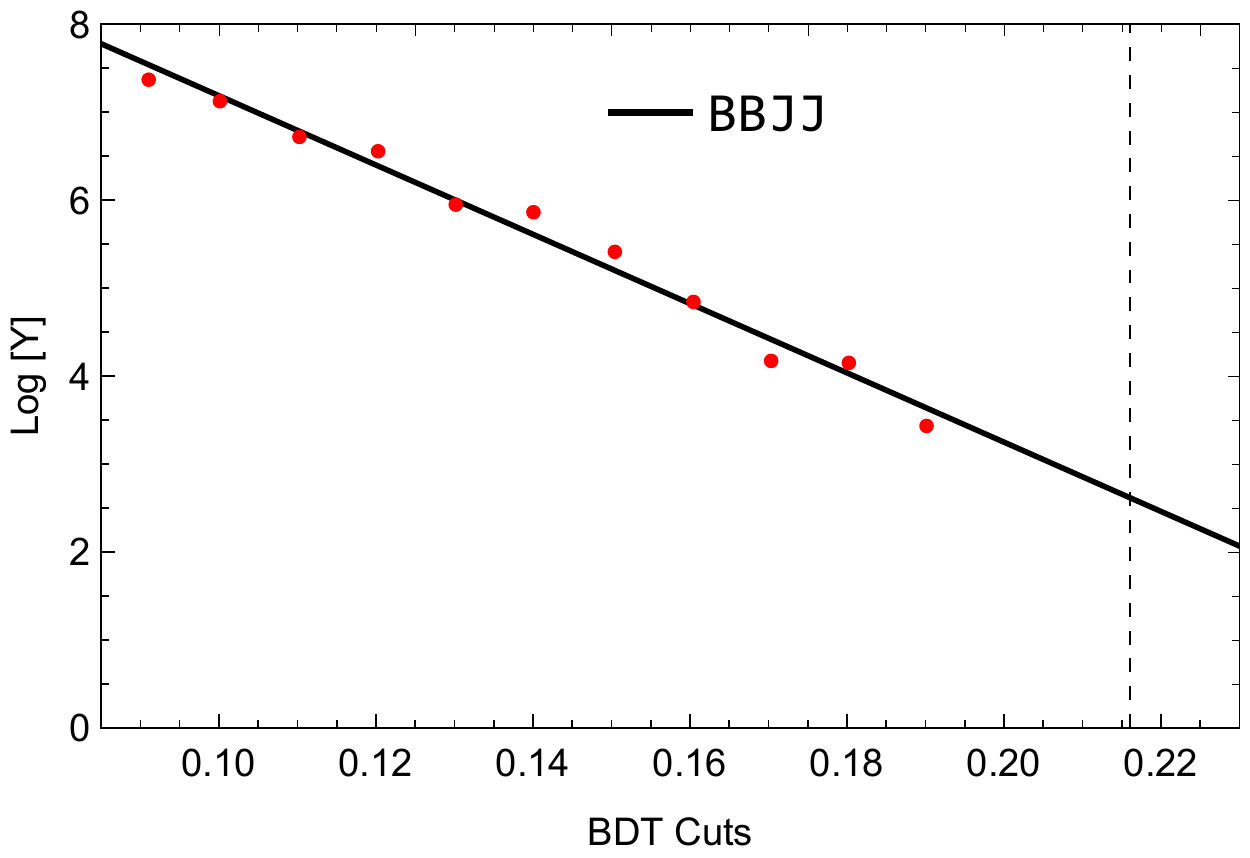}
\includegraphics[width=3.0in]{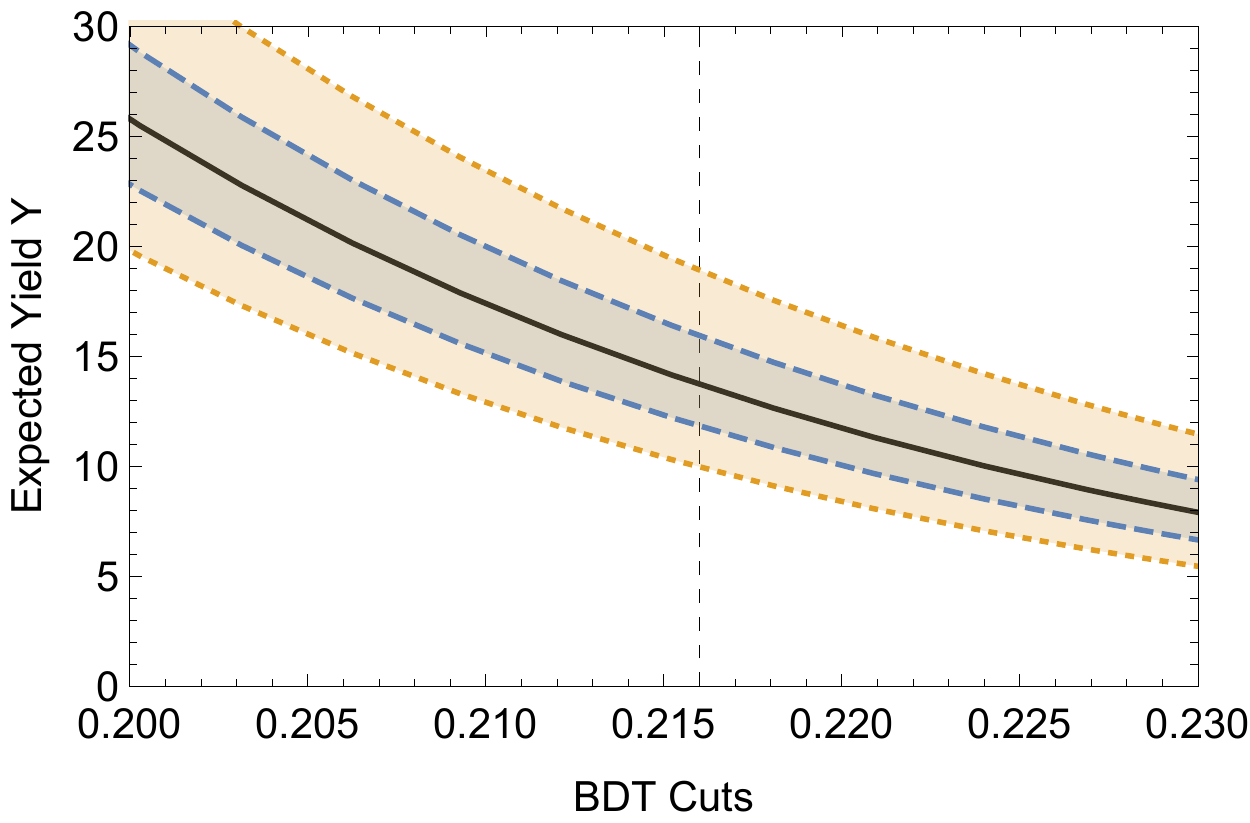}
\caption{Behavior of the $b\bar b jj$ background yield $Y_{b\bar bjj}$ versus
the BDT response cut.
In the upper panels, the MC data points are denoted by bullets and
the solid line in the upper-right panel shows the result of the linear
fitting to $\log Y_{b\bar bjj}$.
In the lower panel, we show the result of extrapolation of the solid line
to the region with BDT Cut $>0.19$, where no data points exist, together with
$1$- and $2$-$\sigma$ errors.
The vertical lines locate the BDT response cut of $0.216$
taken for BDT$_{\rm SM}$.
}
\label{fig:bbjj}
\end{figure}
Precisely, we study the behavior of the $b\bar b jj$ background yield $Y_{b\bar bjj}$ versus
the BDT response cut.
First we observe that, based on the current $b\bar b jj$ MC sample,
our estimation of the background results in $0$ when BDT Cut $>0.19$, see
the upper-left panel of Fig.~\ref{fig:bbjj}.
To extrapolate to the region with BDT Cut $>0.19$, we implement
a linear fitting to $\log Y_{b\bar bjj}$, see the solid line
in the upper-right panel of Fig.~\ref{fig:bbjj}.
And we find that
\begin{equation}
Y_{b\bar bjj}=13.7^{+2.2\,(5.2)}_{-1.9\,(3.7)}
\end{equation}
at $68(95)$\% CL as shown in the lower panel of Fig.~\ref{fig:bbjj}.
Taking the $1\sigma$ upper value of $15.9$,
the number of total background increases by
the amount of about $3$\% which hardly affect
our main results significantly.

Incidentally, we note that  the
$jj\gamma\gamma$ background survives though its cross section
is smaller than that of the $b\bar b\gamma\gamma$ one
by about {\it two} orders of magnitude taking account of
the fake rate $P_{j\to b}$. This is because
its kinematical distributions quite resemble to those of the signal.
For example, compared to other non-resonant backgrounds,
we find that it is quite populated
in the region of $\Delta R_{bb}\lsim 2$ where
most signal events are located.

\setcounter{equation}{0}
\section{Supplemental materials}
\begin{figure}[t!]
\vspace{-1.0cm}
\centering
\includegraphics[width=3.2in]{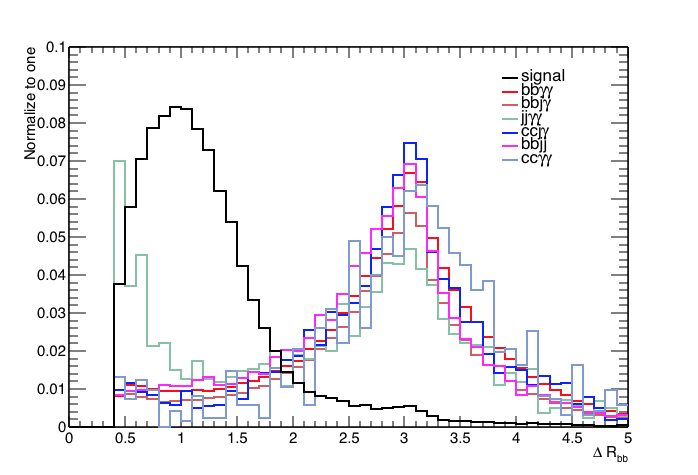}
\includegraphics[width=3.2in]{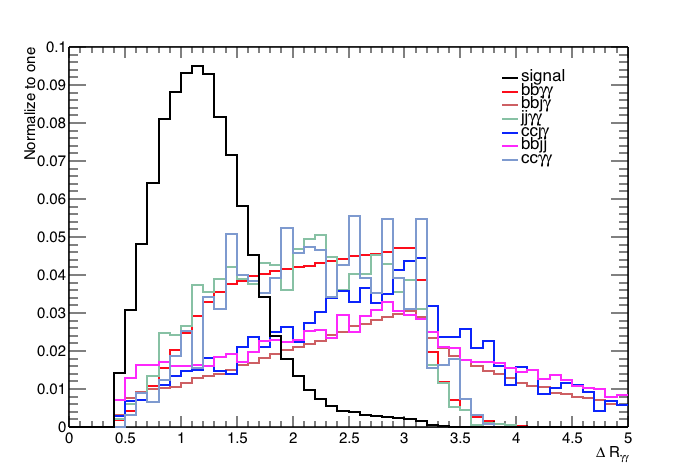}
\includegraphics[width=3.2in]{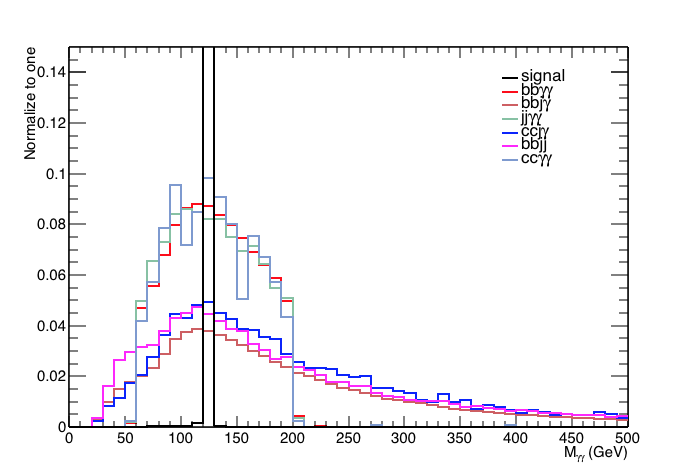}
\includegraphics[width=3.2in]{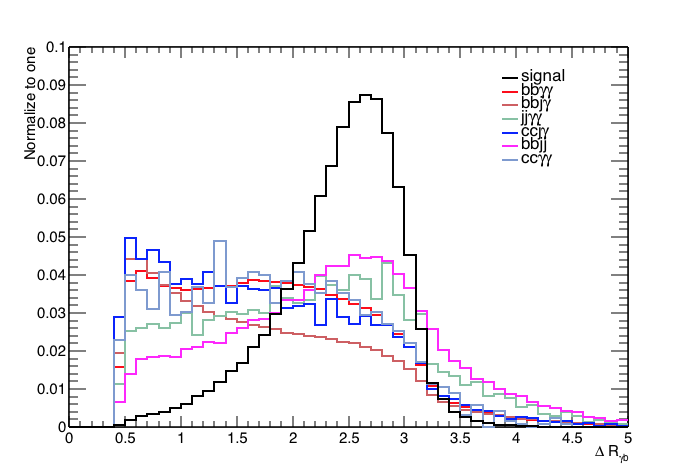}
\includegraphics[width=3.2in]{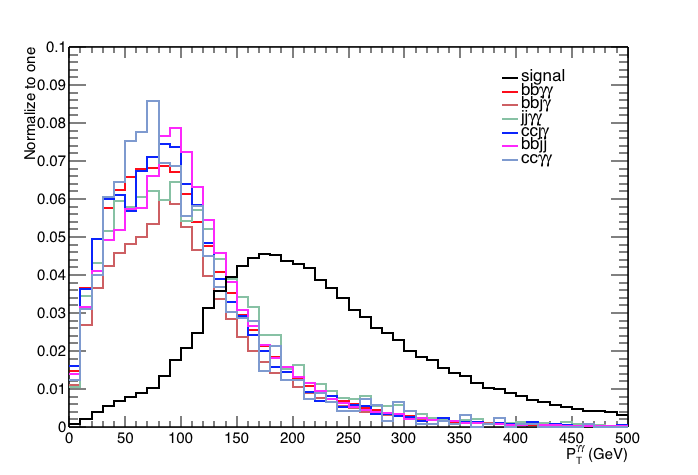}
\includegraphics[width=3.2in]{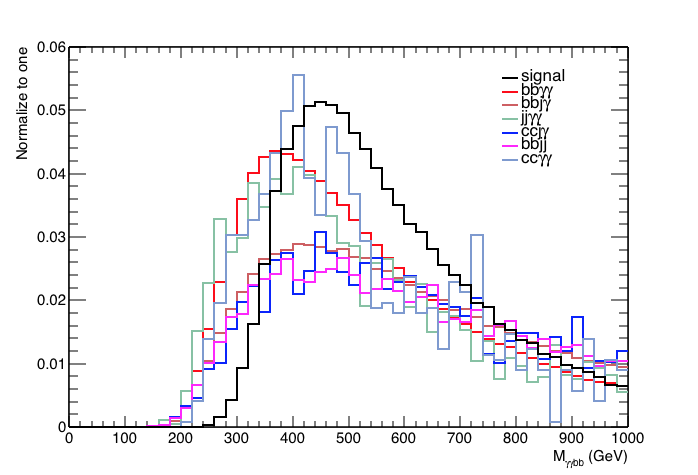}
\includegraphics[width=3.2in]{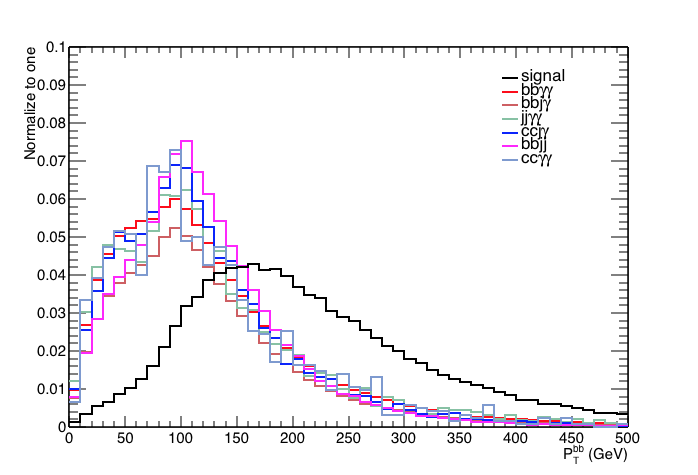}
\includegraphics[width=3.2in]{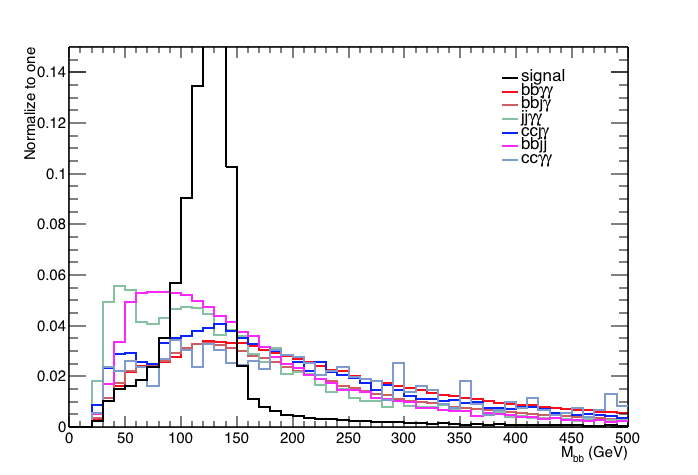}
\caption{Normalized distributions of the eight kinematic variables
for the SM signal and six non-resonant backgrounds after applying the event preselection
cuts 1-5 in Table~\ref{tab:event_selection}.
Panels are in the descending order of importance 
(see Table~\ref{tab:ranking}) from upper-left to lower-right.}
\label{fig:app1}
\end{figure}

In this appendix, we present the
normalized distributions of the eight kinematic variables
for the SM signal with $\lambda_{3H}=1$ (black solid) and 
the six non-resonant backgrounds 
after applying the event preselection cuts 1-5 in Table~\ref{tab:event_selection},
see Fig.~\ref{fig:app1}.
For $M_{bb\,,\gamma\gamma}$ and $P_T^{bb\,,\gamma\gamma}$, in terms of $P_T$,
we choose the least energetic two photons or two $b$ quarks
while the most energetic ones are chosen for $\Delta R_{bb,\gamma\gamma}$ and
$M_{\gamma\gamma bb}$. For $\Delta R_{\gamma b}$, on the other hand, we choose the least
energetic $b$ and the next-to-the-least energetic photon.

%
%
%
\end{appendix}

\bigskip

\newpage

\section*{Acknowledgment}
This work was supported by the National Research Foundation of Korea 
Grant No. NRF-2016R1E1A1A01943297 (J.C., J.S.L., J.P.), 
No. NRF-2018R1D1A1B07051126 (J. P.), 
and by the MoST of Taiwan under Grant No. 107-2112-M-007-029-MY3 (K. C.).


\end{document}